\documentstyle[11pt]{article}%[showkeys]
\newcount\Mac  \Mac=0 % devo mettere Mac=1 se sto lavorando sul file Mac
\newcommand{\lx}{\lambda}
\newcommand{\Lx}{\Lambda}
\newcommand{\ex}{\epsilon}

\newcommand{\sx}{\sigma}

\newcommand{\rht}{\tilde{\rho}}

\newcommand{\kx}{\kappa}

\newcommand{\Gammak}{\Gamma_k}

\newcommand{\be}{\begin{equation}}
\newcommand{\ee}{\end{equation}}
\newcommand{\een}{\end{subequations}}
\newcommand{\ben}{\begin{subequations}}
\newcommand{\beq}{\begin{eqnarray}}
\newcommand{\eeq}{\end{eqnarray}}

\def \lta {\mathrel{\vcenter
     {\hbox{$<$}\nointerlineskip\hbox{$\sim$}}}}
\def \gta {\mathrel{\vcenter
     {\hbox{$>$}\nointerlineskip\hbox{$\sim$}}}}

\def\Red{}
\def\Black{}
\def\Blue{}

\newcommand{\lascia}[1]{}
\def\puttag(#1,#2)#3{\put(#1,#2){\makebox(0,0){\rm\Blue #3\Black}}}

\def\circa#1{\,\raise.3ex\hbox{$#1$\kern-.75em\lower1ex\hbox{$\sim$}}\,}

\def\SO{{\rm SO}}

\newcommand{\riga}[1]{\noalign{\hbox{\parbox{\textwidth}{#1}}}\nonumber}

\def\putps(#1,#2)(#3,#4)#5#6{\ifnum\Mac=1 \put(#1,#2){\special{picture #5}}
\else  \put(#3,#4){\includegraphics{#6}} \fi}

\def\Op{{\cal W}}
\def\One{\hbox{1\kern-.24em I}}
\newcommand{\phibounce}{\phi_{\rm b}}
\newcommand\Ord{{\cal O}}

\makeatletter
\def\art{\@ifnextchar[{\eart}{\oart}}
\def\eart[#1]#2#3#4#5#6{{\rm #2}, {\e, #3 \bf #4} {\rm (#6) #5} ({\em #1})}
\def\hepart[#1]#2{{\rm #2, \em#1}}
\newcommand{\oart}[5]{{\rm #1}, {\em #2 \bf #3} {\rm (#5) #4}}
\newcounter{alphaequation}[equation]
\def\thealphaequation{\theequation\hbox to
0.6em{\hfil\alph{alphaequation}\hfil}}
\def\eqnsystem#1{
\def\@eqnnum{{\rm (\thealphaequation)}}
\def\@@eqncr{\let\@tempa\relax \ifcase\@eqcnt \def\@tempa{& & &} \or
  \def\@tempa{& &}\or \def\@tempa{&}\fi\@tempa
  \if@eqnsw\@eqnnum\refstepcounter{alphaequation}\fi
\global\@eqnswtrue\global\@eqcnt=0\cr}
\refstepcounter{equation} \let\@currentlabel\theequation \def\@tempb{#1}
\ifx\@tempb\empty\else\label{#1}\fi
\refstepcounter{alphaequation}
\let\@currentlabel\thealphaequation
\global\@eqnswtrue\global\@eqcnt=0 \tabskip\@centering\let\\=\@eqncr
$$\halign to \displaywidth\bgroup \@eqnsel\hskip\@centering
$\displaystyle\tabskip\z@{##}$&\global\@eqcnt\@ne
\hskip2\arraycolsep\hfil${##}$\hfil& \global\@eqcnt\tw@\hskip2\arraycolsep
$\displaystyle\tabskip\z@{##}$\hfil
\tabskip\@centering&\llap{##}\tabskip\z@\cr}

\def\endeqnsystem{\@@eqncr\egroup$$\global\@ignoretrue} \makeatother

\begin{document}
\begin{quote}
{\em June 1998}\hfill {\bf SNS-PH/98-12}\\
hep-ph/9806453 \hfill{\bf IFUP-TH/98-24}
\end{quote}
\vspace{2cm}
\begin{center}
{\Large\bf\Red A Consistent Calculation of Bubble-Nucleation Rates}\\[2cm]
\Black\large
{\bf Alessandro Strumia}\\[0.3cm]
\normalsize\em 
Dipartimento di Fisica, Universit\`a di Pisa and\\
INFN, Sezione di Pisa, I-56127 Pisa, Italia\\[0.3cm]
\large {\rm and}\\[3mm] {\bf Nikolaos Tetradis}\\[3mm]
\normalsize\em  Scuola Normale Superiore,\\
Piazza dei Cavalieri 7, I-56126 Pisa, Italia\\[15mm]
\Blue\large\bf Abstract
\end{center}
\begin{quote}\large\indent
We present a consistent picture of tunnelling in field theory.
Our results apply both to high-temperature field theories 
in four dimensions and to zero-temperature three-dimensional ones.
Our approach is based on the notion of a coarse-grained potential
$U_k$ that
incorporates the effect of fluctuations with characteristic momenta 
above a given scale $k$.
$U_k$ is non-convex and becomes equal to the convex
effective potential for $k \to 0$. 
We demonstrate that a consistent calculation of
the nucleation rate must be performed at non-zero values of 
$k$, larger than the 
typical scale of the saddle-point configuration that dominates tunnelling.
The nucleation rate is exponentially suppressed by the action $S_k$ 
of this saddle
point. The pre-exponential factor $A_k$, which includes the
fluctuation determinant around the saddle-point configuration, is well-defined
and finite. Both $S_k$ and $A_k$ are $k$-dependent, but this dependence
cancels in the expression for the nucleation rate.
This picture breaks down in the limit of very weakly first-order phase
transitions, for which the pre-exponential factor compensates 
the exponential suppression.
\end{quote}\Black

\thispagestyle{empty}\newpage\setcounter{page}{1}

\setcounter{equation}{0}
\renewcommand{\theequation}{\thesection.\arabic{equation}}

\section{Introduction}

The consistent description of first-order phase transitions
is a difficult problem which has attracted the attention of
statistical and particle physicists for a long time
(for a review see ref. \cite{review} and references therein).
Our present understanding of these phenomena is based on
the work of Langer on nucleation theory~\cite{langer}.
His formalism has been applied to relativistic field theory by
Coleman~\cite{coleman} and Callan~\cite{colcal} and extended by
Affleck~\cite{affleck} and Linde~\cite{linde}
to finite temperature.
The basic quantity in this approach is the nucleation rate, which
gives the probability per unit time and volume to nucleate a certain
region of the stable phase (the true vacuum) within the metastable 
phase (the false vacuum). 
The rate is exponentially suppressed by the free energy of 
the critical bubble, which is a static configuration
(usually assumed to be spherically symmetric) within the metastable phase 
whose interior consists of the stable phase.
This configuration has a certain radius that can be determined from the
parameters of the underlying theory. Bubbles slightly larger
than the critical one expand rapidly, thus converting the 
metastable phase into the stable one. 
Possible deformations of the critical  bubble
generate a static pre-exponential factor in the nucleation rate.
The leading contribution to this factor  
has the form of a fluctuation determinant. 
Another dynamical prefactor determines the fast growth rate
of the bubbles that are slightly larger than the critical one
\cite{langer,kapusta}. In this work we concentrate on the
calculation of the static prefactor. We are, therefore, mostly 
concerned with the rate of nucleation of critical bubbles.
Their real-time evolution after their nucleation is a separate
question, which we hope to address in a future publication.

The nucleation rate 
per unit volume $I$ 
(probability of nucleation of a critical bubble
per unit time and volume)
for a four-dimensional field theory at temperature
$T$, in the limit that thermal fluctuations dominate over
quantum fluctuations, is given by~\cite{colcal}--\cite{linde}
\be
I=\frac{E_0}{2\pi}
\left(\frac{S}{2\pi }\right)^{3/2}\left|
\frac{\det'[\delta^2 \Gamma/\delta\phi^2]_{\phi=\phibounce}}
{\det[\delta^2 \Gamma/\delta\phi^2]_{\phi=0}}\right|^{-1/2}
\exp\left(-S\right). 
\label{rate0} \ee
Here $\Gamma$ is the free energy of the system for a given configuration of the
field $\phi$ that acts as the order parameter of the problem. 
The rescaled
free energy of the critical bubble is $S=\Gamma_b/T
=\left[\Gamma\left(\phibounce(r)\right)-\Gamma(0)\right]/T$,
where $\phibounce(r)$ is the spherically-symmetric
bubble configuration and $\phi = 0$ corresponds to the false vacuum.
The fluctuation determinants are evaluated either at $\phi = 0$ 
or around $\phi= \phibounce(r)$. 
The prime in the fluctuation determinant around
the bubble denotes that the three zero eigenvalues 
of the operator $[\delta^2 \Gamma/\delta\phi^2]_{\phi=\phibounce}$
have been removed. 
Their contribution generates the factor 
$\left(S/2\pi \right)^{3/2}$ and the volume factor
that is absorbed in the definition of $I$ (nucleation rate per unit volume). 
The quantity $E_0$ is the square root of
the absolute value of the unique negative eigenvalue.
This last contribution appears only for the high-temperature theory
\cite{affleck}. It is absent in the expression for the quantum-tunnelling rate 
in the zero-temperature three-dimensional theory.

The free energy of the critical bubble can be easily determined 
either analytically or numerically. The bubble is the dominant 
saddle-point configuration that interpolates between the two
vacua. Its profile is determined by a differential equation,
which can be integrated numerically or even solved analytically in
simple cases. The calculation of the fluctuation determinants
in the prefactor is a more difficult 
task. They can be brought in a more manageable form if one 
employs spherical coordinates. However, the true difficulty 
concerns the ultraviolet divergences that are inherent in
their calculation. An appropriate regularization scheme must be
employed in order to control them~\cite{cott}--\cite{schmidt}. 

\medskip

The situation becomes even more complicated in the case of 
radiatively induced first-order phase transitions. These are a 
consequence of the appearance of a new vacuum state in the 
theory as a result of the integration of (quantum or thermal) fluctuations
\cite{colwein}.
In field theory the free energy 
(more precisely the thermodynamic
potential) density of a system for homogeneous configurations 
is usually identified with
the temperature-dependent effective potential. A radiatively induced 
first-order phase transition appears in theories for which the 
tree-level potential has only one minimum, while a second minimum
appears at the level of radiative corrections, usually 
computed within a perturbative scheme. This approach, however,
faces two fundamental difficulties:
\begin{itemize}

\item[a)] The effective potential, being the Legendre transform of the
generating functional for the connected Green functions, 
is a convex function of the field. Consequently, it does not 
seem to be the appropriate quantity for the study of tunnelling,
as no structure with more than one minima separated by a barrier
exists\footnote{
It has been argued in ref.~\cite{wu} that the appropriate quantity for
the study of tunnelling is the generating functional of the 
1PI Green functions (calculated perturbatively), 
which differs from the effective potential 
in the non-convex regions. However, as we show in the following,
the consistent picture must rely on the notion of coarse graining
and on the separation of the high-frequency fluctuations that are responsible
for the non-convexity of the potential, from the low-frequency ones
that are relevant for tunnelling. Such notions cannot be easily 
implemented in the context of perturbation theory.}.

\item[b)] The fluctuation determinants in the expression for the nucleation
rate have a form completely analogous to the one-loop correction to
the potential. The question of double-counting the effect of
fluctuations (in the potential and the prefactor)
must be properly addressed. 

\end{itemize}
In this paper we demonstrate that all the above issues can be resolved
through the implemention of the notion of coarse graining in the 
formalism. The appropriate quantity for the description of the
physical system is the {\em effective average action\/}~\cite{averact}, which 
is the generalization in the continuum   
of the blockspin action 
of Kadanoff~\cite{kadanoff}.
The dependence of this action on the coarse-graining scale $k$
is described by an exact flow equation~\cite{exact}
typical of the Wilson approach to the renormalization group
\cite{wilson}. The formalism has been applied with success to second-order
phase transitions. A complete picture has
emerged for the phase transitions in a variety of scalar models,
with a reliable determination of both non-universal quantities
(such as critical temperatures) and universal ones (such
as critical exponents, the equation of state and crossover curves)
\cite{trans}--\cite{twoscalar}. 
The generalization of the formalism to gauge theories 
\cite{gauge} has led to 
the study of second-order and first-order phase transitions for
the Abelian and non-Abelian Higgs models~\cite{abelian4d}--\cite{me},
with implications for
the electroweak phase transition~\cite{me}.
The method has also been applied to the chiral phase transition 
\cite{chiral}.
The framework for the discussion of nucleation rates in first-order phase
transitions has been set in
ref.s~\cite{bubble1,bubble2}\footnote{For a related work see ref. \cite{alford}.}.

\medskip

We first 
summarize the basic notions in
the calculation of nucleation rates in our approach.
We define the bare theory 
at some high scale $\Lx$ that can be identified with the ultraviolet 
cutoff. The renormalized theory at lower scales $k$ is
described in terms of the effective average action
$\Gamma_k$, which can be interpreted as the coarse-grained 
free energy at a given scale $k$. Fluctuations with 
characteristic momenta $q^2 \gta k^2$ are integrated out
and their effect is incorporated in 
$\Gamma_k$. The $k$ dependence 
of $\Gamma_k$ is determined by an exact flow equation. 
This can be translated into evolution equations
for the invariants appearing in a derivative expansion of
the action. We consider only the effective average potential
$U_k$ and a standard kinetic term and neglect higher derivative
terms in the action. The validity of our assumption is discussed 
in the next section. At scales $k$ below the temperature $T$,
the theory can be described in terms of an effective  
three-dimensional action at zero temperature~\cite{trans,me}.
This dimensional reduction indicates the absence of explicit
time dependence for the parameters of the theory. 
It is a consequence of our implicit
assumption that the high-frequency modes of the system are 
in thermal equilibrium and their time dependence
can be averaged out. On the contrary, the low-frequency modes 
have real-time dynamics that are related to the behaviour of the
system during and after nucleation. In this work
we concentrate on static properties of the system, such as the 
characteristics of the critical bubble and 
the nucleation rate. For such quantities, the 
description in terms of an effective three-dimensional theory 
is sufficient. The study of dynamical questions, such as the
expansion rate of bubbles slightly larger than the critical one, 
requires a study of the full four-dimensional theory, at
least for the low-frequency modes. 

We determine the form of the potential $U_k$ at scales
$k$ below the temperature through a numerical solution of
the evolution equation. We are interested in theories for which
$U_k$ has two minima separated by a barrier for low values of $k$.
This structure may exist already at the level of the bare potential,
or appear as the result of the integration of the high-frequency modes.
When $k^2$ becomes smaller than the typical 
positive curvature of the convex parts of $U_k$,
the massive modes that induce the evolution of the potential decouple.
As a result the convex regions of the potential stop evolving.
However, the non-convex part (the barrier) continues to evolve. 
Full convexity is approached in the limit
$k \rightarrow 0$~\cite{convex,largen,analytical}. It is a consequence of
the integration of 
configurations that interpolate between the minima 
in the functional integral that defines $U_k$~\cite{convex, polonyi}. 
All explicit 
information about tunnelling is lost in the resulting Maxwell construction
for the effective potential.

It is clear from the above that the calculation of the 
nucleation rate must be performed at a non-zero value of $k$, such 
that configurations that interpolate between the minima are not taken
into account. This value must be chosen so 
that the convex parts of the potential
have stopped evolving significantly, while a well-defined barrier still
exists.
For a range of values of $k$ that satisfy this requirement we
perform the calculation of the nucleation rate. The profile and 
the free energy of the
bubble are determined in the standard way. The evaluation of the
fluctuation determinants in the prefactor is again performed following
standard techniques, but with an important modification. An
ultraviolet cutoff of order $k$ is imposed, such that fluctuations
with characteristic momenta $q^2 \gta k^2$ are not included. The 
reason is that the effect of such fluctuations is already 
incorporated in $U_k$. This modification resolves two of the serious
problems mentioned earlier. The pre-exponential factor is
now finite and no double-counting of the fluctuations takes place. 

\smallskip

However, an important issue arises at this point. 
The scale $k$ was introduced in the problem as a mere
calculational tool. If our approach makes sense, {\em the choice
of $k$ should not affect physical 
parameters\/} such as the nucleation rate.
The remarkable outcome of our study is that this
expectation is confirmed. Despite the significant $k$ dependence
of the free energy of the bubble and the prefactor, the nucleation
rate is $k$-independent to a good accuracy.  
A residual small $k$ dependence can be interpreted as a measure 
of the contribution of the next order in the saddle-point 
approximation for fluctuations around the bubble.

\medskip

In the following sections we present the details of the calculation
outlined above for a theory of a real scalar field. 
In order not to obscure the essential physics by 
complicating the model too much, we have made some simplifications.
We do not discuss the evolution of $\Gamma_k$ for $k \gta T$. 
For readers who are interested in the details of the
mechanism of dimensional reduction in our approach, detailed
discussions can be found in ref.s~\cite{trans,twoscalar,me}
for a variety of models. We start the 
evolution at a scale $k_0$ 
sufficiently below the temperature of the system, so that the 
dynamics is three-dimensional to a good approximation. 
As an initial condition we consider a potential $U_{k_0}$ with
two inequivalent minima separated by a barrier. 
This form of the potential is determined by the bare potential 
$U_{\Lx}$ and the integration of fluctuations between the scales
$\Lx$ and $k_0$. Some of these fluctuations may correspond to
additional massive degrees of freedom that decoupled above
the scale $k_0$.
In the next section, we describe in detail the initial form of the
potential we use. We integrate
the evolution equation 
for the effective three-dimensional 
theory starting at the scale $k_0$,
and perform the calculation of the nucleation rate
as described above. A significant evolution,
with a substantial variation of the form of the 
potential, may take place
between $k_0$ and the range of scales where we compute
the nucleation rate.

In the following section~we derive the evolution equation for
the potential and describe the initial condtion for its integration.
In section~3 we discuss the formalism we employ 
for the calculation of the nucleation rate. Our results
are presented in section~4 and our conclusions are given 
in section~5.

\setcounter{equation}{0}
\section{Evolution equation for the potential}

In this section~we summarize the formalism of the effective average
action for a theory of a real scalar field $\phi$ and derive the 
evolution equation for the potential. 
We discuss the effective three-dimensional theory that 
results from the dimensional reduction of a high-temperature 
four-dimensional theory at scales below the temperature. 
The temperature can be absorbed in a redefinition of the field and
its potential, so that these have dimensions appropriate
for an effective three-dimensional theory
\beq
\phi & = &\frac{\phi_4}{\sqrt{T}}
\nonumber \\
U(\phi) & = &\frac{U_4(\phi_4,T)}{T}.
\label{fivethree} \eeq
In this way, the temperature does not appear explicitly in
our expressions. This has the additional advantage of 
permitting the straightforward application of our results
to the problem of quantum tunnelling in a three-dimensional 
theory at zero temperature.

\subsection{Intuitive derivation}

Before presenting the rigorous derivation, 
it is instructive to derive the evolution equation for the 
potential based on an intuitive argument, along the lines of
ref.s~\cite{averact,trans}. 
We start by considering the $Z_2$-symmetric scalar model, 
in Euclidean three-dimensional space.
The one-loop effective potential is
given by the expression
\beq
U^{(1)}_k(\rho) &=& 
V(\rho) + \frac{1}{2} 
\ln \det  \left[ P_k + V'(\rho) + 2 V''(\rho) \rho \right]
\nonumber \\
&=& V(\rho) + \frac{1}{2} 
\int_{\Lx} \frac{d^3q}{(2 \pi)^{3}}~
\ln \left[ P_k(q) + V'(\rho) + 2 V''(\rho) \rho \right],
\label{twoone} \eeq
where 
$V(\rho)$ is the bare potential.
In order to be consistent with the conventions in previous
publications, we have defined the variable 
\be
\rho = \frac{1}{2} \phi^2,
\label{twotwo} \ee
which we 
frequently use in this section.
Primes denote derivatives with respect to $\rho$: 
$V'(\rho) =  {d V}/{d \rho}$. 
In terms of this variable, the mass term of the scalar field
is $d^2V/d\phi^2=V'(\rho) + 2 V''(\rho) \rho$. 
The inverse propagator $P_k(q)$ in momentum space for a massless field
is given by 
$P_{k=0}(q)=q^2$ in perturbation theory. 
We assume that the momentum integration is regulated by an ultraviolet
cutoff $\Lx$.

We would like to introduce an effective infrared cutoff $k$ for the 
low-frequency modes, so that the momentum integration in
eq.~(\ref{twoone}) does not receive contributions from modes with 
characteristic momenta $q^2 \lta k^2$.
The simplest way to achieve this is through
the addition of a mass term $k^2$ to the 
perturbative inverse propagator, so that for a massless field
\be
P_k(q) = q^2 + k^2. 
\label{twothree} \ee
The potential
now depends on $k$, as indicated by the subscript in eq.~(\ref{twoone}). 

The next step is to derive an evolution equation for the change of 
$U_k$ with the scale $k$ and follow the evolution for 
$k \rightarrow 0$.  
For this purpose we take the logarithmic derivative
with respect to $k$ 
and substitute $U_k$ for $V$ 
in the right-hand side of eq.~(\ref{twoone}). The
intuitive justification for this replacement is based on the fact 
that the new contributions to the momentum integration, when $k$ is
lowered by a small amount $\Delta k$, come from the region 
$k- \Delta k < q < k$. The relevant mass term and couplings that should
appear in the evolution equation are the renormalized ones at the 
scale $k$ (which, for the scalar field, are related to derivatives of $U_k$)
and not the bare ones. This ``renormalization-group improvement''
results in the evolution equation 
\beq
\frac{\partial U'_k(\rho)}{\partial t} &=& -\frac{1}{2} 
\int \frac{d^3q}{(2 \pi)^{3}}~ \frac{\partial P_k}{\partial t}
\frac{3U''_k(\rho)
+2 U'''_k(\rho)\rho}{\left[ 
P_k(q) + U_k'(\rho) + 2 U_k''(\rho) \rho \right]^2} 
\label{ttt} \\
&=& - \frac{k^2}{8 \pi}~ \frac{3U''_k(\rho)+2 U'''_k(\rho)\rho}{\sqrt{k^2+
U_k'(\rho) + 2 U_k''(\rho) \rho}},
\label{twofour} \eeq
where
$t = \ln (k/\Lx)$, with $\Lx$ identified with
the ultraviolet cutoff of the theory.
We have derived the evolution equation for 
$U'_k(\rho)$ (and not $U_k(\rho)$) because this is the easiest to 
integrate numerically. 
For $k=\Lx$ the infrared and ultraviolet cutoffs coincide, and no
integration of fluctuations takes place. This determines the 
initial conditions for the solution of eq.~(\ref{twofour})
as $U_{\Lx}(\rho) = V(\rho)$.
In the opposite limit, $k \rightarrow 0$, one 
recovers the effective potential $U(\rho) \equiv U_0(\rho)$.

One could look for an iterative solution of eq.~(\ref{ttt}).
The first iteration results in the equation
\be
U^{(1)}(\rho) = 
U_k(\rho) + \frac{1}{2} 
\ln  \frac{ \det \left[  P_{k=0} + U_k'(\rho) + 2 U_k''(\rho) \rho \right]
}{ \det \left[ P_k + U_k'(\rho) + 2 U_k''(\rho) \rho \right]}
\label{btwoone} \ee
for the effective potential. This expression can be compared with
eq.~(\ref{twoone}) with $k=0$. The two equations have the same
structure, but the bare potential $V(\rho)$ is replaced by the 
$k$-dependent potential $U_k(\rho)$.
Also, the determinant resulting from the radiative corrections 
is replaced by a ratio of determinants.
The numerator of this ratio is what one would expect from eq.~(\ref{twoone})
with $k=0$. The denominator involves the inverse propagator $P_k$ 
and effectively removes the fluctuations with characteristic momenta
$q^2 \gta k^2$. This is justified by the fact that the effect of these 
fluctuations has already been incorporated in $U_k(\rho)$.
There is no need for the addition of an extra regulator $\Lx$ as 
in eq.~(\ref{twoone}). Its role is played by $k$, which acts
as an effective ultraviolet cutoff in the calculation of 
the effective potential from $U_k(\rho)$.

The above observation will be important in the following 
section~where nucleation rates will be computed. 
The first part of 
our procedure will be to integrate the evolution equation
(\ref{ttt}) numerically, 
from a scale $k_0$, where we choose the initial form of the
potential $U_{k_0}(\rho)$,
down to a non-zero scale $k$. 
This effective integration of the high-frequency modes generates
the appropriate potential that describes the dynamics of low-frequency modes
with $q^2 \lta k^2$.
Tunnelling will be discussed in the context of the 
low-energy theory through the standard saddle-point approximation
\cite{langer,coleman}.
The first correction to the leading semiclassical result involves
the usual ``one-loop'' form of a fluctuation determinant
$\det \left[ P_{k=0} + {\rm {\cal O}} \right]$,
where ${\rm {\cal O}}$ is an operator to be defined in the next section.
A consistent treatment of the effect of fluctuations, which avoids 
double-counting the high-frequency modes, can be obtained 
if this correction is modified to
$\det \left[\left( P_{k=0} + {\rm {\cal O}}\right)/
\left( P_k + {\rm {\cal O}}\right) \right]$
as in eq.~(\ref{btwoone}).

\subsection{Rigorous derivation}

The evolution equation for the potential can be derived within a 
more rigorous
approach through the formalism of the effective average action~\cite{averact,
exact,indices}.
The 
effective average action $\Gamma_k$, 
for a theory described by a bare action $S$,  
results from the effective integration
of degrees of freedom with characteristic momenta larger than a given 
infrared cutoff $k$. Its dependence 
on the scale $k$ is described by an exact flow equation.
In this subsection we summarize the formalism
for the case of a $Z_2$-symmetric theory of a real scalar field
in Euclidean three-dimensional space. 
A detailed discussion can be found in 
ref.s~\cite{exact,indices}. 

We specify the action together with some ultraviolet 
cutoff $\Lambda$, so that the theory is properly 
regulated. 
We add to the kinetic term a piece that has the
following form in momentum space 
\be
\Delta_k S [\chi] = \frac{1}{2} \int d^3\!q~
R_k(q) \chi^*(q) \chi(q),
\label{twoonea} \ee
where $\chi^*(q)=\chi(-q)$.
The function $R_k$  is used 
to prevent the propagation of modes 
with characteristic momenta $q^2 \lta k^2$. 
As a result, the inverse propagator 
for the action $S + \Delta S$ has a minimum $\sim k^2$. 
The modes with $q^2 \gg k^2$ are unaffected,
while the 
low-frequency modes with $q^2 \ll k^2$ are cut off:
\be
\lim_{q^2 \rightarrow 0} R_k \sim k^2.
\label{twofoura} \ee
We emphasize at this point that 
many alternative choices of $R_k$ are possible.

We subsequently introduce sources and  
define the generating functional for the connected Green functions 
for the action $S + \Delta S$. Through a Legendre 
transformation we obtain the
generating functional for the 1PI Green functions 
${\tilde \Gamma}_k[\phi]$, where $\phi$ is the expectation value of the 
field $\chi$ in the presence of sources.
The use of the modified propagator for the calculation of 
${\tilde \Gamma}_k$ results in the effective integration of only the 
fluctuations with $q^2 \gta 
k^2$. Finally, the 
effective average action is 
obtained by removing the infrared cutoff 
\be
\Gamma_k[\phi] = {\tilde \Gamma}_k[\phi] -
\frac{1}{2} \int d^3 q
R_k(q) \phi^*(q) \phi(q).
\label{twofive} \ee

For $k$ equal to the ultraviolet cutoff $\Lambda$~\footnote{
For scales $k\gta T$ the high-temperature behaviour of the
four-dimensional theory is relevant. For this reason our
discussion in terms of an effective three-dimensional theory is
not sufficient. In this work, however, we never discuss the evolution
at such high scales, but work at scales $k \leq k_0 \lta T$ instead. 
We refer to the ultraviolet cutoff $\Lx$ only at this point for
reasons of completeness of the presentation. For a full discussion
of the evolution at large momentum 
scales and the mechanism of dimensional reduction,
see ref.s~\cite{trans,twoscalar,me}.
}, 
$\Gammak$ becomes 
equal 
to the bare action $S$ (no effective integration of modes takes 
place), while for $k \rightarrow 0$ it tends towards the effective action 
$\Gamma$ corresponding to $S$ (all the modes are included).
The interpolation of $\Gammak$ between the bare and the 
effective action makes it a very useful field-theoretical tool.
The means for practical calculations is provided by an exact 
flow equation\footnote{
For other versions of exact renormalization group equations, 
see ref.s~\cite{wilson,reneq}.},
which describes the response of the 
effective average action to variations of the infrared cutoff 
($t=\ln(k/\Lambda)$)~\cite{exact}:
\be
\frac{\partial}{\partial t} \Gammak[\phi]
= \frac{1}{2} {\rm Tr} \left\{ (\Gammak^{(2)}[\phi] + R_k)^{-1} 
\frac{\partial}{\partial t} R_k \right\}. 
 \label{twosix} \ee 
Here $\Gammak^{(2)}$ is the second 
functional derivative of 
$\Gamma_k$
with respect to $\phi$. 

Making use of the $Z_2$ symmetry, 
we parametrize the  
effective average action as 
\be
\Gammak = 
\int d^3x \left\{ U_k(\rho) 
+ \frac{1}{2} \partial^{\mu} \phi~{:} Z_k(\rho,-\partial^2) {:}~
\partial_{\mu} \phi~  +\cdots
\right\},
\label{twoeleven} \ee
where the normal ordering indicates that the derivative operators are
always on the right. 
The dots stand for invariants that involve more
derivatives of the field.
In order to turn the flow equation for the 
effective average action into equations for 
$U_k$, $Z_k$, etc, 
we have to evaluate the trace in eq.~(\ref{twosix}) 
for properly chosen 
background field configurations.
For the evolution equation for $U_k$ we have to expand around a constant 
field configuration. We find~\cite{exact,indices}
\be
\frac{\partial}{\partial t} U_k(\rho) = 
\frac{1}{2} \int \frac{d^3\!q}{(2 \pi)^3}~ 
\frac{\partial R_k(q)/\partial t }{Z_k(\rho,q^2)  q^2 
+ R_k(q) + U'_k(\rho) + 2 \rho U''_k(\rho)}.
\label{threefour} \ee

The above equation is an exact evolution equation for the
potential. Its solution, however, requires 
information on the wave-function renormalization $Z_k(\rho,q^2)$.
Throughout this paper we shall set
$Z_k(\rho,q^2)=1$, which corresponds to the 
first order of the derivative expansion of eq.~(\ref{twoeleven}).
This is expected to be a good approximation because the 
size of $Z_k$ is related to the anomalous dimension of the field,
which is small for this model ($\eta \simeq 0.035$ for the three-dimensional
theory we consider).  
For $\eta=0$ the kinetic term in the
$k$-dependent inverse propagator must be exactly proportional to 
$q^2$ both for $q^2 \rightarrow 0$ and 
$q^2 \rightarrow \infty$.
Several studies have confirmed the smallness of the 
corrections arising from the deviation of 
$Z_k(\rho,q^2)$ from 1~\cite{indices,num,eos}.
Within our approximation, eq.~(\ref{threefour}) 
reproduces eq.~(\ref{ttt}) through the definition
\be
P_k(q) = q^2 + R_k(q).
\label{fiftyone} \ee
For the choice
\be
R_k(q) = k^2,
\label{fiftytwo} \ee
eq.~(\ref{twofour}) is obtained. 

We should point out that the choice of eq.~(\ref{fiftytwo})
for the cutoff function $R_k(q)$ may be problematic. For example, 
in four dimensions it results in divergent integrals in the right-hand side
of evolution equations such as eq.~(\ref{threefour}). Even for the
three-dimensional case that we are considering, the right-hand side
of eq.~(\ref{threefour}) involves an irrelevant 
$\rho$-independent divergent constant, which disappears in the 
evolution equation for $U'_k(\rho)$. For these reasons, in most
cases it is preferable to work with a cutoff function of the 
form 
\be
R_k(q) = \frac{q^2~\exp\left(-{q^2}/{k^2} 
\right)}{1-\exp\left(-{q^2}/{k^2} \right)},
\label{fiftyfive} \ee
for which integrals such as the one in the right-hand side of
eq.~(\ref{threefour}) are finite. 
However, such a cutoff function corresponds to a non-local operator
in position space, which would make the calculations of the
next section impossible. Because of this, we employ the 
cutoff function of eq.~(\ref{fiftytwo})
in this paper and emphasize that care must be taken
in the extension of our discussion to the case of full four-dimensional
dynamics.

\subsection{Scale-invariant form of the evolution equation}

It is convenient to cast the evolution equation (\ref{twofour}) in 
a form that does not explicitly depend on the scale $k$.
This makes the identification of possible fixed points easier.
For this reason we define the dimensionless quantities
\beq
u_k(\rht) &=&~\frac{U_k(\rho)}{k^3}
\nonumber \\
\rht &=&~\frac{\rho}{k}.
\label{athreeone} \eeq
Primes on $u_k$ denote derivatives with respect to $\rht$.
We can now rewrite the evolution equation for the potential 
as
\be
\frac{\partial u'_k}{\partial t} = 
~-2 u'_k + \rht u''_k
+\frac{1}{8 \pi^2} 
(3 u''_k +2 \rht u'''_k) L^3_1 \left( u'_k +2 u''_k \rht \right).
\label{athreethree} \ee
The non-trivial solution of the above equation with 
${\partial u'_k}/{\partial t}=0$ corresponds to the 
Wilson-Fisher fixed point that determines the dynamics of
the second-order phase transition in the 
$Z_2$-symmetric theory~\cite{trans,indices,eos}. 

The dimensionless function $L^3_1(w)$ 
is given by 
\beq
L^3_1(w) &=&- \frac{1}{2\pi k}
\int d^3\!q~\frac{\partial P_k}{ \partial t} (P_k + w)^{-2} \nonumber \\
&=&
- \frac{1}{k}
\int_0^{\infty} dx~\sqrt{x}
\frac{\partial P_k}{ \partial t} (P_k + w)^{-2},
\label{athreefour} \eeq
with $x=q^2$.
This has been discussed extensively in ref.s~\cite{averact,indices,convex} 
(for various forms of the infrared-regulating function $R_k$ and in various
dimensions).
It has the interesting property that
it falls off for large values of $w$ following a power law. As 
a result it introduces a threshold behaviour for the 
contributions of massive modes to the evolution equation. 
The third term in the right-hand side of 
the evolution equation (\ref{athreethree})
includes the $L^3_1$ function with the mass of the $\phi$ field 
divided by $k^2$ as its argument.
When the scale $k^2$ crosses below the running squared 
mass this
contribution vanishes and the massive mode decouples. 
As a result the evolution of $U_k(\rho)$ stops. 
The function $L^3_1(w)$ also has a pole at $w=-1$. This property 
induces the convexity of the potential in the limit $k \to 0$.
The argument of $L^3_1(w)$ in the non-convex regions is given 
by the negative curvature of the potential divided by $k^2$.
As the pole cannot be crossed, the curvature follows $k$ to zero,
thus inducing the convexity of the effective potential.  
For the choice of eq.~(\ref{fiftytwo})
for the cutoff function, we obtain
\be
L^3_1(w) = -\frac{\pi}{\sqrt{1+w}},
\label{fiftyeight} \ee
in agreement with eq.~(\ref{twofour}).

Two algorithms for the numerical integration of eq.~(\ref{athreethree})
have been presented in detail in ref.~\cite{num}.
The comparison of the two methods 
provides a good check on possible systematic numerical 
errors. The two algorithms give results that
agree at the 0.3\% level. We expect 
the numerical solution to be an approximation of the solution
of the partial differential equation (\ref{athreethree}) 
with the same level of accuracy. 

\subsection{Explicit breaking of the $Z_2$ symmetry}

Up to this point we have not discussed the breaking of the 
$Z_2$ symmetry that could lead to vacuum instability. 
The formalism of the previous subsections is completely
$Z_2$-invariant and has to be modified in order to account for
the symmetry-breaking effects. However, we shall work within
a framework for which the modifications are minimal.
As we explained in the introduction, we consider an effective
three-dimensional theory that results from the dimensional
reduction of a high-temperature four-dimensional one. 
We define this theory at a scale $k_0 \lta T$. 
The form of $U_k$ is determined by the bare potential 
$U_{\Lx}$ and the integration of fluctuations between the scales
$\Lx$ and $k_0$. Some of these fluctuations may correspond to
additional massive degrees of freedom that decoupled above
the scale $k_0$. We choose a form of the potential that breaks the 
$Z_2$ symmetry, while permitting the presence of a non-trivial
evolution at scales $k\leq k_0$.

Following ref.~\cite{seide}, we consider theories that are 
described by potentials of the form
\be
U_{k_0} (\phi) = 
\frac{1}{2}m^2_{k_0} \phi^2
+\frac{1}{6} \gamma_{k_0} \phi^3
+\frac{1}{8} \lx_{k_0} \phi^4.
\label{two20} \ee
By a variable shift 
\be
\sx = \phi + \frac{\gamma_{k_0}}{3 \lx_{k_0}}
\label{two21} \ee
we can bring this potential to the form 
\begin{eqnsystem}{sys:Ugamma}
U_{k_0} (\sx) &=& c_{k_0} - J_{\gamma} \sx
+ \frac{1}{2}\mu^2_{k_0} \sx^2
+\frac{1}{8} \lx_{k_0} \sx^4,
\label{two22} \\
\riga{with}\\[-2mm]  
J_{\gamma} &=& \frac{\gamma_{k_0}}{3 \lx_{k_0}} m^2_{k_0}
-\frac{\gamma^3_{k_0}}{27 \lx^2_{k_0}}, \\
\mu^2_{k_0} &=& m^2_{k_0}
-\frac{\gamma^2_{k_0}}{6 \lx_{k_0}}.
\label{two23}
\end{eqnsystem}
The right-hand side of the 
exact flow equation for the effective average action 
of the $\sx$ field is not affected by the linear term 
$- J_{\gamma} \sx$ or the constant $c_{k_0}$. 
Therefore, 
the evolution equation for the potential can be integrated
using the $Z_2$-symmetric
formalism. The potential of $\phi$ is recovered through
the relation 
\be
U_k(\phi) = c_{k_0} - J_{\gamma} \sx+U_k^{Z_2}(\sx)
=
c_{k_0} - J_{\gamma}\cdot \left(
\phi + \frac{\gamma_{k_0}}{3 \lx_{k_0}} \right)
+U_k^{Z_2} \left( \phi + \frac{\gamma_{k_0}}{3 \lx_{k_0}} \right),
\label{two24} \ee
where $U_k^{Z_2}$ is the potential of the $Z_2$-symmetric model
with mass term $ \mu^2_{k_0}$ and quartic coupling $\lx_{k_0}$.
In ref.~\cite{seide} it has been verified that the above procedure
gives the same result as the straightforward integration of
the evolution equation~(\ref{athreethree}) with an initial condition given by
eq.~(\ref{two20}).

In the following we integrate numerically eq.~(\ref{athreethree})
for $Z_2$-symmetric models. Different types 
of evolution can be obtained by keeping $\lx_{k_0}$ constant and
varying the mass term $ \mu^2_{k_0}$. More specifically, 
the Wilson-Fisher 
fixed point of the three-dimensional theory can be approached for
a certain (negative) 
value $( \mu^2_{k_0})_{cr}$. This fixed point
determines the properties of the second-order phase transition
of the $Z_2$-symmetric theory. 
Through the
appropriate choice of $\gamma_{k_0}$, we can vary the difference in
the energy density between the minima of the potential 
according to eq.~(\ref{two24}). Arbitrarily weakly first-order phase
transitions can be studied in the limit $\mu^2_{k_0} \rightarrow 
( \mu^2_{k_0})_{cr}$, $\gamma_{k_0} \rightarrow 0$.

\setcounter{equation}{0}
\section{Calculation of the nucleation rate}

\subsection{General formalism}

As we discussed in the previous sections, we study tunnelling in 
a theory of a real scalar field with 
a Euclidean action
\be
\Gammak = 
\int d^3x \left\{ 
\frac{1}{2}(\partial^{\mu} \phi)( \partial_{\mu} \phi )
+ U_k(\phi) \right\}
\label{ctwoeleven} \ee
and a coarse-graining scale $k \not= 0$.
The parameters in the above action are effective three-dimensional
ones defined according to eqs.~(\ref{fivethree}).
The potential $U_k(\phi)$ has two minima corresponding to
vacua with different vacuum energy densities:
the stable (true)
minimum is located at $\phi=\phi_t$ and the unstable (false) one at
$\phi=\phi_f=0$.

The unrenormalized decay rate per unit volume 
from the false minimum towards the true one is given by 
\cite{coleman,colcal,cott}
\be
I=\frac{E_0}{2\pi}
\left(\frac{S_k}{2\pi}\right)^{3/2}\left|
\frac{\det'[\delta^2 \Gamma_k/\delta\phi^2]_{\phi=\phibounce}}
{\det[\delta^2 \Gamma_k/\delta\phi^2]_{\phi=0}}\right|^{-1/2}
\exp({-S_k}). 
\label{rate} \ee
This is analogous to eq.~(\ref{rate0}) after the absorption 
of the explicit factors of $T$ in the redefinition of the field
and potential and the introduction of a coarse-graining scale. 
Notice that if the factor 
${E_0}/{2\pi}$ is removed, the above expression reproduces the 
quantum-tunnelling rate of the zero-temperature three-dimensional theory 
\cite{colcal,affleck}. 
As this factor gives only a small contribution to the total rate, our
discussion applies to quantum tunnelling as well within a good approximation.

The nucleation rate is exponentially suppressed by 
the action $S_k$ (the rescaled free energy) of the bubble configuration
$\phibounce(r)$. This is an 
$\SO(3)$-invariant solution of
the classical equations of motion which interpolates between 
the local maxima of the potential
$-U_k(\phi)$. It satisfies the equation 
\be
{d^2\phibounce\over dr^2}+\frac{2}{r}~{d\phibounce\over dr}=U_k'(\phibounce), 
\label{eom} \ee
with the boundary conditions 
$\phibounce\rightarrow 0$ for $r \rightarrow \infty$ and
$d\phibounce/dr= 0$ for $r =0$.
The action $S_k$ of the bubble is given by 
\be
S_k=4 \pi
\int_0^\infty 
\left[ \frac{1}{2}
\left(\frac{d\phibounce(r)}{dr}\right)^2
+U_k(\phibounce(r)) \right]
r^{2}\,dr
\equiv  S_k^t+S_k^v,
\label{action} \ee 
where the kinetic and potential contributions, 
$S_k^t$ and $S_k^v$ respectively,
satisfy 
\be
{S_k^v\over S_k^t}=-{1\over 3}.
\label{const1} \ee

The pre-exponential factor 
corresponds to the first correction to the semiclassical approximation
in the saddle-point method. 
The numerator is the 
fluctuation determinant 
around the bubble
\be
{\det}'\left[ \delta^2 \Gamma_k/\delta\phi^2 \right]_{\phi=\phibounce}=
{\det}'\left[-\partial^2+U''_k(\phi=\phibounce(r))
\right],
\label{fluct1} \ee
while the denominator is
the fluctuation determinant
around the false vacuum $\phi=\phi_f=0$
\be
\det\left[ \delta^2 \Gamma_k/\delta\phi^2 \right]_{\phi=0}=
\det\left[ -\partial^2+ 
U''_k(\phi=0)\right].
\label{fluct2} \ee
In this section we revert to the standard way of denoting
the derivatives with respect to $\phi$ with primes. This should not
be confusing as we always indicate 
the quantity with respect to which
we differentiate as the argument of the function.
The differential operator $-\partial^2+U''_k
(\phibounce(r))$ has three 
zero modes (the three spatial translations of the bounce).
The prime over the determinant indicates that 
these modes have to be omitted 
in its calculation. 
Their contribution generates the factor $(S_k/2\pi)^{3/2}$ in 
eq.~(\ref{rate}) and the volume factor
that is absorbed in the definition of $I$ (nucleation rate per unit volume).
The quantity $E_0$ is the square root of
the absolute value of the unique negative eigenvalue of the above operator.

The pre-exponential factor defined in eq.~(\ref{rate})
is in general ultraviolet-divergent and an appropriate regularization
scheme must be employed. 
Within our approach, the form of the regularization 
is dictated by the discussion at the end of subsection~2.1.
The effect of the high-frequency modes has been incorporated in the 
form of the coarse-grained potential $U_k(\phi)$, which is obtained
through the integration of the evolution equation (\ref{athreethree}). 
In order not to double-count this effect, fluctuation
determinants computed within the low-energy theory must be 
replaced by a ratio of determinants, in complete analogy to 
eq.~(\ref{btwoone}). This implies that a consistent expression for
the nucleation rate is given by 
\beq
I
&\equiv&A_k \exp({-S_k})
\nonumber \\
A_k&=& \frac{E_0}{2\pi}\left(\frac{S_k}{2\pi}\right)^{3/2}
\times
\nonumber \\
&~&
\left|
\frac{\det'\left[-\partial^2+
U''_k(\phibounce(r))
 \right]}
{\det \left[ -\partial^2+R_k\left(-\partial^2 \right) +
U''_k(\phibounce(r))
\right]}
~\frac{\det\left[-\partial^2+R_k\left(-\partial^2 
\right)+
U''_k(0)
\right]}
{\det\left[
-\partial^2+
U''_k(0)
\right]}
\right|^{-1/2}.
\label{rrate} \eeq
The infrared-regulating function $R_k$ has been discussed in the
previous section (see eq.~(\ref{fiftyone})). The simplest
choice for this function, which is necessary for the feasibility of 
the computation of
the next section, is given by eq.~(\ref{fiftytwo}), i.e.
$R_k=k^2$. We point out that only the operator 
$-\partial^2+U''_k(\phibounce(r))$
has negative and zero eigenvalues that
require special treatment in eq.~(\ref{rrate}).

\begin{figure}[t]
\begin{center}\hspace{-5mm}
\begin{picture}(17,11)
\putps(0,-0.4)(-2.8,-0.5){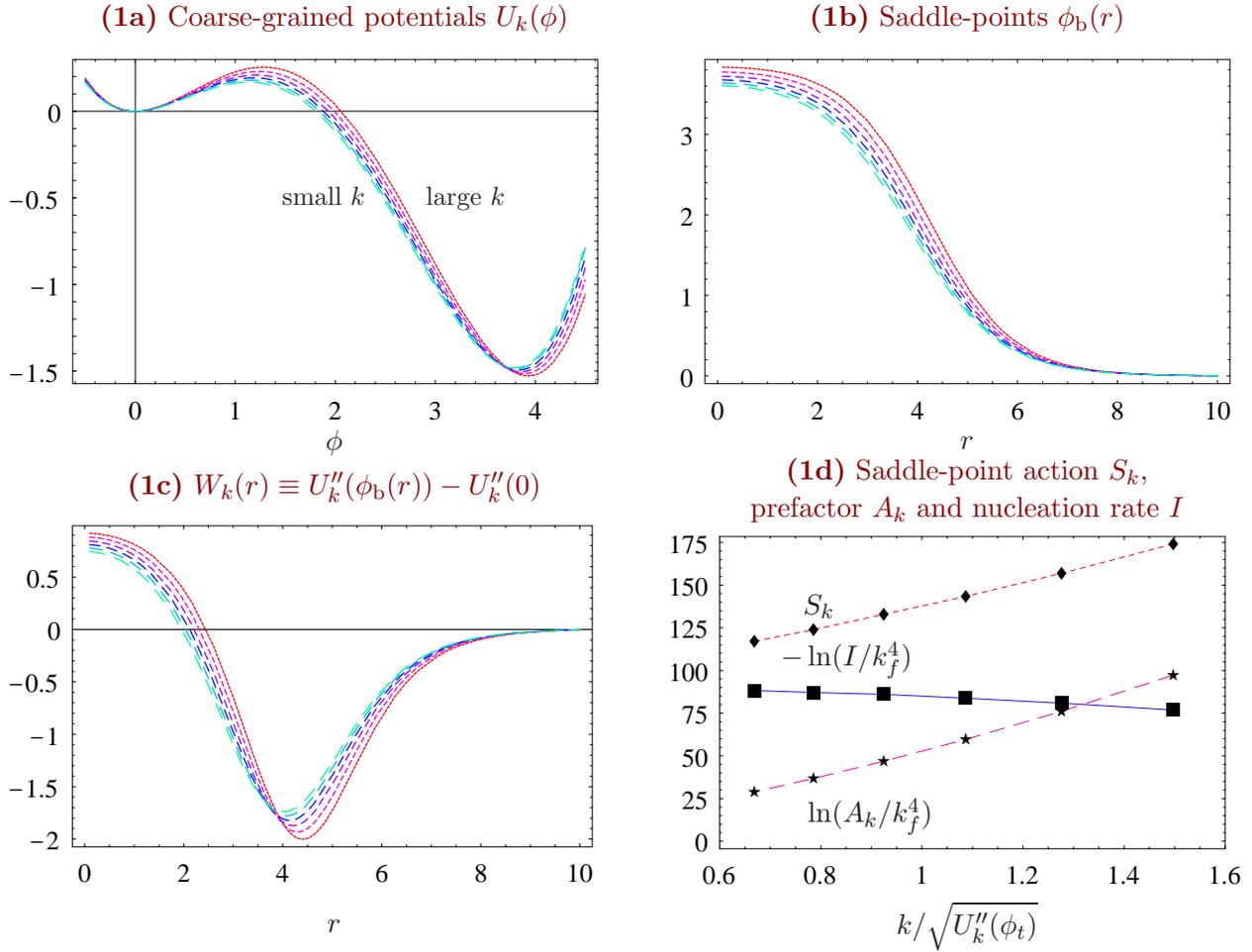}{figEx.ps}\Red
\put(5,11.5){\makebox(0,0){{\bf (1a)} Coarse-grained potentials $U_k(\phi)$}}
\put(13.5,11.5){\makebox(0,0){{\bf (1b)} Saddle-points $\phibounce(r)$}}
\put(5,5.2){\makebox(0,0){{\bf (1c)} $W_k(r)\equiv U''_k(\phibounce(r))-U''_k(0)$}}
\put(13.5,5.4){\makebox(0,0){{\bf (1d)} Saddle-point action $S_k$,}}
\put(13.5,4.9){\makebox(0,0){prefactor $A_k$ and nucleation rate $I$}}
\Black
\put(5,5.8){\makebox(0,0){$\phi$}}
\put(13.5,5.8){\makebox(0,0){$r$}}
\put(5,-0.7){\makebox(0,0){$r$}}
\put(13.5,-0.7){\makebox(0,0){$k/\sqrt{U_k''(\phi_t)}$}}
\put(4.3,9){\small small $k\qquad$ large $k$}
\put(11,3.5){$\phantom{-}S_k$}
\put(11,2.8){$-\ln(I/k_f^4)$}
\put(11,0.7){$\phantom{-}\ln (A_k/k_f^4)$}
\end{picture}
\vspace{1cm}

\caption[SP]{\em The steps in the computation of the nucleation rate 
for a model with
$\mu^2_{k_0}/ k_0^2=-0.05$, $\lambda_{k_0}/k_0=0.1 $, 
$\gamma_{k_0}/k_0^{3/2}=-0.0634$.
The dimensionful quantities are given in units of $k_f=0.223~k_0$.
\label{fig:Ex}}
\end{center}\end{figure}

\subsection{Details of the numerical computation}

From this point on, we 
use the form $R_k=k^2$ for the infrared-regulating function.
The profile of the bubble can be easily computed 
with the ``shooting'' method~\cite{shooting}. We integrate eq.
(\ref{eom}) numerically, starting at $r=0$ with a value of $\phi$ near
the true minimum $\phi_t$ and $d\phi/dr=0.$ We then adjust the initial value
of $\phi$ so that the boundary condition 
$\phibounce\rightarrow 0$ for $r \rightarrow \infty$ is satisfied. 
(In practice this condition is satisfied at a value $r_\infty$
sufficiently larger than the typical size of the bubble.)

The computation of the fluctuation determinants is more complicated. 
The differential operators that appear in eq.~(\ref{rrate}) 
have the general form
\begin{eqnsystem}{sys:W}
\Op_{\kappa\alpha}&=&-\partial^2 +m_\kappa^2+\alpha W_k(r), \label{op} \\
\riga{where}\\[-2mm]
m_\kappa^2&\equiv&U''_k(0)+\kappa k^2,\\
W_k(r)&\equiv&U''_k(\phibounce(r))-U''_k(0),\label{ak}
\end{eqnsystem}
with $\kx,\alpha=0$ or 1.
Since the $\Op_{\kappa\alpha}$ operators are SO(3) symmetric, it is convenient
to use spherical coordinates $(r,\theta,\varphi)$
and express the eigenfunctions $\psi$ in terms of spherical harmonics:
$\psi(r,\theta,\varphi)=u(r)/r~Y_{\ell m}(\theta,\varphi)$~\cite{cott,baacke1}.
Here $\ell$ and $m$ are the usual angular quantum numbers.
The Laplacian operator $\partial^2$ takes the form
\be
-\partial^2~~~\to~~~ \frac{1}{r} 
\left[-\frac{d^2}{dr^2}+\frac{\ell(\ell+1)}{r^2}\right]r
\equiv -\frac{1}{r}\nabla^2_\ell r,
\label{sph} \ee
so that
\beq
\det \Op_{\kappa\alpha}
&=&\prod_{\ell=0}^\infty (\det \Op_{\ell\kappa\alpha})^{2\ell+1}
\nonumber \\
\Op_{\ell\kappa\alpha}&=&-\nabla^2_\ell+m_\kappa^2+\alpha W_k(r).
\label{opl} \eeq
We recall that 
$\det \Op_{\ell \kx \alpha}$ is defined as the product of all eigenvalues 
$\lambda$ that lead to solutions of 
${\cal W}_{\ell \kx\alpha} u(r)=\lambda u(r)$, 
with the function $u(r)$ vanishing at $r=0$ and $r \to \infty$.
The computation of such complicated determinants is made possible by 
a powerful theorem~\cite{erice,cott} that relates ratios
of determinants to solutions of ordinary differential equations.
In particular, we have  
\be
g_{\ell\kappa}\equiv
\frac{\det \Op_{\ell \kappa 1}}{\det \Op_{\ell \kappa 0}}
=\frac{\det[-\nabla^2_\ell+m_\kappa^2+1\cdot W_k(r)]}
{\det[-\nabla^2_\ell+m_\kappa^2+0\cdot W_k(r)]}=
\frac{y_{\ell\kappa1}(r\to \infty)}{y_{\ell\kappa0}(r \to \infty)},
\label{theorem} \ee
where $y_{\ell\kappa\alpha}(r)$ is the solution of the differential equation
\be
\left[-\frac{d^2}{dr^2}+\frac{\ell(\ell+1)}{r^2}
+m_\kappa^2+\alpha W_k(r)\right]y_{\ell\kappa\alpha}(r)=0,
\label{diffeq} \ee
with the behaviour
$y_{\ell\kappa\alpha}(r)\propto r^{\ell+1}$ for $r\to 0$.
Such equations can be easily solved numerically with Mathematica 
\cite{Mathematica}.
In the ``free case'' ($\alpha=0$) it is possible to obtain the exact analytical
solution\footnote{The function $i_n$ is the standard Bessel 
$I$ function, defined as
$i_{n}(z)={\tt BesselI}[n,z]$ in Mathematica notation~\cite{Mathematica}.
For $n=\ell+1/2$ a semi-integer, 
$i_n$ can be expressed in terms of elementary functions as
$\sqrt{z}i_{\ell+1/2}(z)=P_+^\ell(1/z)e^z+P_{-}^\ell(1/z) e^{-z}$,
where $P^\ell_{\pm}$ are polynomials of 
degree $\ell$. For $z\to\infty$ (or, more precisely, for $z\gg \ell$)
$i_{\ell+1/2}(z)\to e^z/\sqrt{2\pi z}$.}
\be
y_{\ell\kappa0}\propto i_{\ell+1/2}(m_\kappa r)/\sqrt{m_\kappa r}.
\label{sol0} \ee
The final expression for the nucleation rate, 
appropriate for an efficient numerical computation, is
\beq
I &=& \frac{1}{2 \pi}
\left(\frac{S_k}{2\pi}\right)^{3/2}\exp\left(-S_k\right)
c_0 c_1
\prod_{\ell=2}^\infty c_\ell,
\nonumber \\
c_0 &=& \left( \frac{E_0^2 g_{01}}{\left| g_{00} \right|}
\right)^{1/2},~~~~~~~~~~
c_1 = \left({g_{11}\over g'_{10}}\right)^{3/2},~~~~~~~~~~
c_{\ell} = \left({g_{\ell 1}\over g_{\ell 0}}\right)^{(2\ell+1)/2}.
\label{fin} \eeq
The factors $c_\ell$ for $\ell \geq 2$ can be computed in a straightforward 
way through eqs.~(\ref{theorem}), (\ref{diffeq}) as we explained above.
The calculation of $c_1$ is more complicated because of the necessity to
eliminate the zero eigenvalues in $g'_{10}$. This can be achieved by 
replacing the operator $\Op_{10\alpha}$ with 
$\Op_{10\alpha}+\ex~U''_k(0)$
and evaluating $g'_{10}$ as~\cite{cott}
\be
g'_{10}=
\frac{1}{U''_k(0)}~ \lim_{\ex \rightarrow 0}
\left[ \frac{1}{\ex}~\frac{\det \Op_{1 0 1}(\ex)}{\det \Op_{1 0 0}} \right].
\label{g1} \ee
The (unique) negative 
eigenvalue $-E_0^2$ 
of $\Op_{001}$ can be obtained by solving the 
equation $\Op_{001}u=-E_0^2 u$ and using the shooting method
to determine the value of $E_0^2$ that
ensures the correct boundary condition $u(r \rightarrow \infty)=0$.

As a final remark, we give the explicit expression for $c_{\ell}$
in the limit of large $\ell$. 
It can be 
obtained by solving the differential equations (\ref{diffeq}), 
using first-order perturbation theory in $W_k$~\cite{cott}.
In terms of 
\begin{eqnsystem}{sys:}
w_n&\equiv&-\int_0^\infty r^n~W_k(r)\,dr,\label{ouf} \\
\riga{we find}\\[-2mm]
g_{\ell0}^{(2\ell+1)/2},\,g_{\ell 1}^{(2\ell+1)/2} &\to& \exp\left(-w_1/2\right)
\left[ 1+\Ord(\ell^{-2}) \right], \\
c_\ell &\to& 1+k^2 \frac{w_3}{4\ell^2}+\Ord(\ell^{-4}).\label{asympt}
\end{eqnsystem}
We have checked that our numerical solution reproduces the
above behaviour for large $\ell$.
The expression (\ref{fin}) for the nucleation rate is finite, as can be
easily checked by considering the identity
\be
\prod_{\ell=2}^\infty 
\left(1+{D^2\over\ell^2}\right)=\frac{\sinh D\pi}{D\pi(1+D^2)}.
\label{infsum} \ee
This last expression is very useful for the numerical computation.
To obtain an accurate result we need to compute 
the exact value of $c_{\ell}$ only up
to $\ell = $10--100. 
Large $\ell>30$ are necessary only for large values of $k$ and/or 
large couplings in the potential.

\newcommand\putTre[4]{\put(3.3,#1){\makebox(0,0){$#2$}}
\put(9,#1){\makebox(0,0){$#3$}}
\put(14.6,#1){\makebox(0,0){$#4$}}}
\begin{figure}
\begin{center}
\begin{picture}(18,21)
\putps(0,-0.5)(-3,-0.5){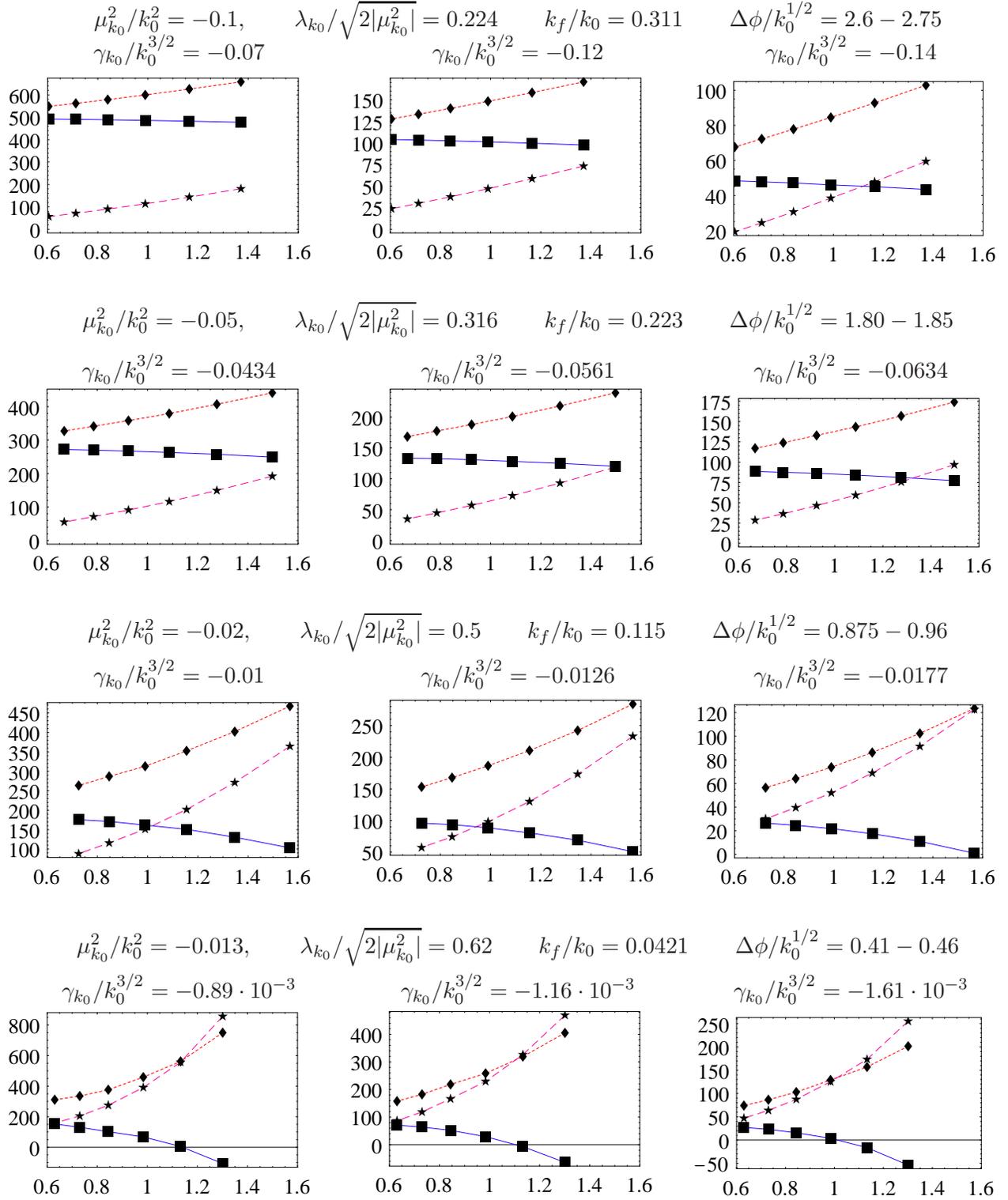 scaled 900}{figRes.ps}
\putTre{19.1}{\gamma_{k_0}/k_0^{3/2}=-0.07}{\gamma_{k_0}/k_0^{3/2}=-0.12}{\gamma_{k_0}/k_0^{3/2}=-0.14}
\putTre{13.733}{\gamma_{k_0}/k_0^{3/2}=-0.0434}{\gamma_{k_0}/k_0^{3/2}=-0.0561}{\gamma_{k_0}/k_0^{3/2}=-0.0634}
\putTre{8.6}{\gamma_{k_0}/k_0^{3/2}=-0.01}{\gamma_{k_0}/k_0^{3/2}=-0.0126}{\gamma_{k_0}/k_0^{3/2}=-0.0177}
\putTre{3.3}{\gamma_{k_0}/k_0^{3/2}=-0.89\cdot 10^{-3}}{\gamma_{k_0}/k_0^{3/2}=-1.16\cdot10^{-3}}{\gamma_{k_0}/k_0^{3/2}=-1.61\cdot10^{-3}}
\put(9,19.6){\makebox(0,0){$\mu^2_{k_0}/k_0^2=-0.1,\qquad \lambda_{k_0}/\sqrt{2|\mu_{k_0}^2|}=0.224\qquad
k_f/k_0=0.311\qquad \Delta\phi/k_0^{1/2}=2.6-2.75$}}
\put(9,14.533){\makebox(0,0){$\mu^2_{k_0}/k_0^2=-0.05,\qquad \lambda_{k_0}/\sqrt{2|\mu_{k_0}^2|}=0.316\qquad
k_f/k_0=0.223\qquad \Delta\phi/k_0^{1/2}=1.80-1.85$}}
\put(9,9.3){\makebox(0,0){$\mu^2_{k_0}/k_0^2=-0.02,\qquad \lambda_{k_0}/\sqrt{2|\mu_{k_0}^2|}=0.5\qquad
k_f/k_0=0.115\qquad \Delta\phi/k_0^{1/2}=0.875-0.96$}}
\put(9,4.0){\makebox(0,0){$\mu^2_{k_0}/k_0^2=-0.013,\qquad \lambda_{k_0}/\sqrt{2|\mu_{k_0}^2|}=0.62\qquad
k_f/k_0=0.0421\qquad \Delta\phi/k_0^{1/2}=0.41-0.46$}}
\end{picture}
\vspace{0.5cm}

\caption[SP]{\em The behaviour of the nucleation rate $I$ for several values of the parameters of the model
with $\lambda_{k_0}	/k_0=0.1$.
We show the values of $S_k$ (diamonds), $\ln (A_k/k_f^4)$ (stars) and $-\ln(I/k_f^4)$
(squares)
as a function of $k/\sqrt{U_k''(\phi_t)}$.
%Inclusion of $A$ makes the $k$ dependence $\{7,5,5/2,2\}$ times milder in the four cases, independently on $\gamma_{k_0}$.
\label{fig:Res}}
\end{center}\end{figure}

\setcounter{equation}{0}\section{Results}
The various steps in our calculation are summarized in fig.~1,
for a theory described by the potential of eqs.~(\ref{two20}), (\ref{two23}) 
with 
$\mu^2_{k_0}=-5 \cdot 10^{-2}~k_0^2$,
$\gamma_{k_0}=-6.34 \cdot 10^{-2}~k_0^{3/2}$,
$\lx_{k_0}=0.1~k_0$.
The dimensionful scale $k_0$ is related to the temperature $T$ of
the system. It is determined by our assumption 
that the effective three-dimensional
description becomes valid at $k_0$. Previous studies of 
dimensional reduction in the context of the effective average action
\cite{trans,me} indicate that $k_0 \simeq T$. 
It must be pointed out at this point that the two-minimum structure is 
often a consequence of the integration of fluctuations of effective 
three-dimensional degrees of freedom (additional scalar or gauge fields).
In such cases the scale $k_0$ must be taken sufficiently small for this
structure to emerge~\cite{me}--\cite{bubble1}\footnote{The 
calculation of the pre-exponential factor must
take into account fluctuations associated with these additional
degrees of freedom~\cite{buch}.}.

In fig.~1a we present the evolution of the potential $U_k(\phi)$ as the scale
$k$ is lowered. We have shifted the metastable vacuum to
$\phi=0$. 
The solid line corresponds to $k/k_0=0.513,$ while the
line with longest dashes (that has the smallest barrier height)
corresponds to $k_f/k_0=0.223$. At the scale $k_f$ the negative 
curvature at the top of the barrier is slightly larger than 
$-k_f^2$. This means that the pole of the function
$L^3_1(w)$ at $w=-1$ is approached in eq.~(\ref{athreethree}).
This is the point in the evolution of the potential
where configurations that 
interpolate between the minima start becoming relevant
in the functional integral that defines the coarse-grained potential
\cite{convex,polonyi}. 
For this reason,
we stop the evolution at this point. 
The potential and the field have been
normalized with respect to $k_f$, so that they are of order 1. 
We observe that, as $k$ is lowered, the absolute minimum of the potential
settles at a non-zero value of $\phi$, while a significant barrier
separates it from the metastable minimum at $\phi=0$. 
The profile of the critical bubble $\phibounce(r)$
is plotted in fig.~1b in units of $k_f$
for the same sequence of scales.  For $k\simeq k_f$ the characteristic 
length scale of the bubble profile and $1/k$ are comparable. This is expected, 
because the form of the profile is determined by the barrier of the potential,
whose curvature is $\simeq -k^2$ at this point. 
This is an additional indication that we should not proceed to coarse-graining
scales below $k_f$.
We observe a significant 
variation of the value of the field $\phi$ in the interior of the bubble
for different $k$.
This is reflected in the form of the quantity $W_k(r)$, defined in
eq.~(\ref{ak}), which we plot in fig.~1c.

\smallskip

Our results for the nucleation rate are presented in fig.~1d.
The horizontal axis corresponds to $k/\sqrt{U''_k(\phi_t})$,
i.e. the ratio of the scale $k$
to the square root of the positive curvature of the potential at the 
absolute minimum. 
The latter quantity gives the mass of the field
at the absolute minimum. 
Typically, when $k$ crosses below this mass (corresponding to
the value 1 on the horizontal axis) the massive fluctuations of the field
start decoupling and the evolution of the convex parts of
the  potential slows down and eventually stops.
The dark diamonds give the values of the action $S_k$ 
of the critical bubble at the scale $k$. We observe a strong 
$k$ dependence of this quantity, which is expected from 
the behaviour in fig.s~1a--1c.
The stars in fig.~1d indicate the values of 
$\ln ( A_k/k^4_f )$. 
Again a strong $k$ dependence is observed. More specifically,
the value of $A_k$ decreases for decreasing $k$. This is expected,
because $k$ acts as the effective ultraviolet cutoff in the calculation 
of the fluctuation determinants in $A_k$. For smaller $k$,
fewer fluctuations with wavelengths above an increasing length scale
$\sim 1/k$ contribute explicitly to the fluctuation determinants. 
The dark squares give our results for 
$-\ln(I/k^4_f ) 
= S_k-\ln ( A_k/k^4_f )$. It is remarkable that the 
$k$ dependence of this quantity disappears as $k$ crosses 
below $\sqrt{U''_k(\phi_t})$ and approaches $k_f$. 
The small residual dependence on $k$ can be used to estimate the 
contribution of the next order in the expansion around the saddle point.
It is reassuring that this contribution is expected to be smaller than
$\ln ( A_k/k^4_f )$. 

This behaviour confirms our expectation that the 
nucleation rate should be independent of the scale $k$ that 
we introduced as a calculational tool. It also demonstrates that
all the configurations plotted in fig.~1b give equivalent 
descriptions of the system, at least for the lower values of $k$.
The implication is that the critical bubble should not be identified
just with the saddle point of the semiclassical approximation, whose
action is scale dependent. It is the combination of 
the saddle point and its possible deformations
in the thermal bath that has physical meaning. 

\medskip

In fig.~2 we present the calculation of the nucleation rate for 
several values of the parameters of the model. 
Each row is computed for initial potentials 
$U_{k_0}$ with the same values of
$\mu^2_{k_0}$ and $\lx_{k_0}$. It also has the same value of
$k_f$. Thus, it corresponds to the same $Z_2$-symmetric theory in the
limit $\gamma_{k_0} = 0$. Different values of 
$\gamma_{k_0}$ are used for the three calculations in each row, so that the
saddle points have different profiles and, therefore, different  
nucleation rates are predicted. 
Moving down the sequence of rows, the value of 
$\lx_{k_0}$ is kept fixed, while 
$| \mu^2_{k_0} |$ is reduced. 
The effective dimensionless coupling 
$\lx_{k_0}/ \sqrt{ 2 | \mu^2_{k_0} |}$
of the $Z_2$-symmetric theory increases, and 
the resulting potentials
have more pronounced barriers relative to the location of the 
minima. This indicates that 
the effect of fluctuations should be enhanced.
The last row corresponds to a $Z_2$-symmetric theory that
starts approaching the Wilson-Fisher fixed point during the
evolution of the potential, before deviating
towards the phase with symmetry breaking or the symmetric one
\cite{indices,eos,me,num}.
We have tried to keep the values of the predicted nucleation
rates comparable in each column of fig.~2.
However, the discontinuity in the field expectation value during
the first-order phase transition $\Delta \phi = \phi_t$ 
decreases as we move down each column. Thus, the strength of
the phase transition diminishes.

The most striking aspect of the comparison of the results in
each column concerns the relative values of 
$S_k$ and $\ln ( A_k/k^4_f )$.
In the first column the contribution of the prefactor to the
nucleation rate is much smaller than that of the action of the 
saddle point. The main role of the prefactor is 
to remove the $k$-dependence from $I/k^4_f$. 
As we move down each column, the difference between 
$S_k$ and $\ln ( A_k/k^4_f )$ diminishes.
In the last row the two quanitities are comparable. 
This confirms our expectation that the effects of fluctuations
should be enhanced in more weakly first-order phase transitions.
The second observation concerns the $k$-dependence of the
predicted nucleation rate. In the last row the contribution from
the prefactor fails to cancel completely the $k$ dependence
of the action of the saddle point. In more quantitative terms, 
when the prefactor is taken
into account the $k$ dependence of the nucleation rate 
is reduced by a factor $\sim 10$ in the first row, while only by 
a factor $\sim 2$ in the last one. 
The reason for the above behaviour is clear. In the last row 
and for $k=k_f$, the nucleation rate is roughly equal to or smaller 
than the contribution from the prefactor. Thus, the effect of 
the next order in the expansion around the saddle point is 
important and can no longer be neglected.
This indicates that there is a limit for the validity of Langer's picture 
of homogeneous nucleation~\cite{langer}. For sufficiently weakly first-order 
phase transitions the saddle point of the semiclassical approximation
is overwhelmed
by the fluctuations around it. As a result, one can no longer
rely on a picture based on the semiclassical approximation.

\setcounter{equation}{0}
\section{Summary and conclusions}

In this paper we addressed the problem of the calculation of nucleation
rates for first-order phase transitions
in the context of high-temperature field theories.
The most commonly employed field-theoretical tool in such
investigations is the effective potential, usually computed
within a perturbative scheme. A first-order phase transition
is expected if two minima are present in the effective potential,
separated by a barrier. 
The nucleation rate is calculated through an expansion
around the saddle point of the functional integral that
interpolates between the two minima.
The rate is given by an expression that involves an exponential
suppression by the action of the saddle point, and a pre-exponential
factor that includes the fluctuation determinant around the saddle-point 
configuration. 

Several major obstacles must be overcome 
before a consistent description can be obtained.
The most obvious one is a consequence of the convexity of
the effective potential. This precludes the discussion 
of tunnelling, as no barrier exists between the minima.
All relevant information 
is washed out by the Maxwell construction. 
In most studies, a perturbative approximation to the 
generating functional of the 1PI Green functions 
is used instead of the effective potential. Such
a quantity has non-convex parts, but it also has 
imaginary parts that are difficult to 
interpret. For radiatively induced first-order phase transitions
(a common occurrence in field theory) there are more 
conceptual difficulties. It is not clear which 
fluctuations of the system generate the two-minimum structure
of the potential and which are associated with the pre-exponential
factor in the calculation of the nucleation rate. It is difficult to
resolve in a clear way the issue 
of double-counting the effect of fluctuations. 
At the technical level, a serious issue concerns 
the ultraviolet 
divergences that appear 
in the calculation of the pre-exponential factor. 
They must be cancelled by counterterms
of the original action, in a way consistent with the calculation
of the potential that determines the action of the saddle point.

In this work we followed an approach that resolves all the
above issues. It relies on the introduction of a coarse-graining
scale $k$ in the problem. This scale separates the high-frequency 
fluctuations of the system, which may be responsible for the 
presence of the second minimum through the Coleman-Weinberg mechanism,
from the low-frequency ones which are relevant for tunnelling. 
The appropriate tool for the calculation of the nucleation rate is
the coarse-grained free energy at a non-zero value of $k$. 
The pre-exponential factor is well-defined and finite,
as the scale $k$ acts as an ultraviolet cutoff in the calculation of
the fluctuation determinants. This is a natural consequence of the fact 
that all fluctuations with typical momenta above $k$ are
already incorporated in the form of the coarse-grained free energy. 

We employed the formalism of the effective average action 
$\Gamma_k$~\cite{averact,
exact,indices}, which we 
identified with the coarse-grained free energy.
As a starting point, we considered 
$\Gamma_k$ for a real scalar field
at a scale $k_0$ below the temperature, such that
the theory has an effective three-dimensional description.
We approximated $\Gamma_k$ by a standard kinetic term and
a potential with two minima 
given by eq.~(\ref{two20}). We assumed that this form of the
potential results from the bare potential 
$U_{\Lx}$ after the integration of (quantum and thermal)
fluctuations between the scales
$\Lx$ and $k_0$. Some of these fluctuations may correspond to
additional massive degrees of freedom that decoupled above
the scale $k_0$. 
If the two-minimum structure is 
a consequence of the integration of fluctuations of effective 
three-dimensional degrees of freedom (additional scalar or gauge fields)
the scale $k_0$ must be taken sufficiently small for this
structure to emerge~\cite{me}--\cite{bubble2}.

We computed the form of the potential $U_k$ at scales $k\leq k_0$ by
integrating an evolution equation derived from an exact flow 
equation for $\Gamma_k$.  
$U_k$ is non-convex for non-zero $k$, and 
approaches convexity only in the limit $k\to 0$.
The nucleation rate must be computed for $k$ larger than the scale $k_f$
at which the functional integral in the definition of $U_k$ starts 
receiving contributions from field configurations that interpolate between
the two minima. This happens when $-k^2$ becomes approximately equal to 
the negative curvature at the top
of the barrier~\cite{convex}. 
For $k > k_f$ the typical length scale of a thick-wall critical
bubble is $\gta  1/k$.

We performed the calculation of the nucleation rate for a range of scales
above and near $k_f$. We found that the saddle-point configuration 
has an action $S_k$ with a significant $k$ dependence. 
For strongly first-order phase transitions, the nucleation
rate $I = A_k \exp(-S_k)$
is dominated by the exponential suppression.
The main role of the prefactor $A_k$, which is also $k$-dependent,  
is to remove the scale dependence from the total nucleation rate.
The implication of our results 
is that the critical bubble should not be identified
just with the saddle point of the semiclassical approximation.
It is the combination of 
the saddle point and its possible deformations
in the thermal bath (accounted for by the fluctuation 
determinant in the prefactor) that has physical meaning. 

For progressively more weakly first-order phase transitions,
the difference between 
$S_k$ and $\ln ( A_k/k^4_f )$ diminishes.
This indicates that the effects of fluctuations
become more and more enhanced. 
At the same time a significant $k$-dependence of the
predicted nucleation rate develops. 
The reason for the above deficiency is clear. 
When the nucleation rate is roughly equal to or smaller 
than the contribution from the prefactor, the effect of 
the next order in the expansion around the saddle point is 
important and can no longer be neglected.
This indicates that there is a limit for the validity of Langer's picture 
of homogeneous nucleation~\cite{langer}. 
For sufficiently weakly first-order 
phase transitions the saddle point of the semiclassical approximation
is dominated by the fluctuations around it. Despite the presence
of a discontinuity in the order parameter, one can no longer 
rely on a picture based on the semiclassical approximation.
An alternative picture must be developed for the description
of the physical system~\cite{gleiser}.

Finally, we point out that our results are relevant for the 
question of quantum tunnelling  in the zero-temperature 
three-dimensional theory.
Only a small numerical factor differentiates between our expressions 
and the ones that determine the quantum-tunnelling rate in that case
\cite{colcal,affleck}. The essential qualitative conclusions remain 
unaffected.

\paragraph{Acknowledgements} We  would like to thank
C. Wetterich for many helpful discussions.
The work of N.T. was supported by the E.C. under TMR contract 
No. ERBFMRX--CT96--0090.

\appendix
\setcounter{equation}{0}
\renewcommand{\theequation}{\thesection.\arabic{equation}}

\end{document}